\documentclass[conference]{IEEEtran}
\usepackage{cite}
\usepackage{geometry}
\newgeometry{left=1.55cm,right=1.55cm,bottom=1.7cm,top=1.7cm}%
\usepackage{threeparttable}
\usepackage{graphicx}
\usepackage{picinpar}
\usepackage[cmex10]{amsmath}
\usepackage{amsmath,amsfonts,amssymb}
\usepackage{subfigure}
\usepackage{algorithm}
\usepackage{algorithmic}
\usepackage{stfloats}
\usepackage{bm}
\usepackage{amsthm}

\begin{document}
\newtheorem{lemma}{Lemma}
\newtheorem{corol}{Corollary}
\newtheorem{theorem}{Theorem}
\newtheorem{proposition}{Proposition}
\newtheorem{problem}{Problem}
\newtheorem{definition}{Definition}
\newcommand{\e}{\begin{equation}}
\newcommand{\ee}{\end{equation}}
\newcommand{\eqn}{\begin{eqnarray}}
\newcommand{\eeqn}{\end{eqnarray}}

\title{Channel Estimation for mmWave Massive MIMO Based Access and Backhaul in Ultra-Dense Network}
%
\author{ \authorblockN{Zhen Gao, Linglong Dai, and Zhaocheng Wang}\\
\authorblockA{Tsinghua National Laboratory for Information
   Science and Technology (TNList),\\
    Department of Electronic Engineering, Tsinghua University, Beijing 100084, P. R. China\\
       Emails: gaozhen010375@foxmail.com}
\thanks{This work was supported in part by the International Science \& Technology Cooperation Program of China (Grant No. 2015DFG12760),
the National Natural Science Foundation of China (Grant Nos. 61571270 and 61271266),  the Beijing Natural Science Foundation (Grant No. 4142027),
and the Foundation of Shenzhen government.}
}
\maketitle
\begin{abstract}
 Millimeter-wave (mmWave) massive MIMO used for access and backhaul in ultra-dense network (UDN) has been considered
 as the promising 5G technique. We consider 
 such an heterogeneous network (HetNet) that ultra-dense small base stations (BSs) exploit
 mmWave massive MIMO for access and backhaul, while macrocell BS provides the control service with low frequency band.
 However, the channel estimation for mmWave massive MIMO can be challenging,
 since the pilot overhead to acquire the channels associated with
 a large number of antennas in mmWave massive MIMO can be prohibitively high.
 This paper proposes a structured compressive sensing (SCS)-based channel estimation scheme,
 where the angular sparsity of mmWave channels is exploited to reduce the
required pilot overhead.
Specifically, since the path loss for non-line-of-sight paths is much larger than that for line-of-sight paths,
the mmWave massive channels in the angular domain appear the obvious sparsity.
 By exploiting such sparsity, the required pilot overhead only depends on the small number of dominated multipath.
 Moreover, the sparsity within the system bandwidth is almost unchanged, which can be exploited for the further improved performance.
Simulation results demonstrate that the proposed scheme outperforms its counterpart, and it can approach
 the performance bound.


\end{abstract}


\IEEEpeerreviewmaketitle

\section{Introduction}\label{S1}
\begin{figure}[tp!]
\vspace*{-1mm}
\centering
\includegraphics[width=0.75\columnwidth, keepaspectratio]{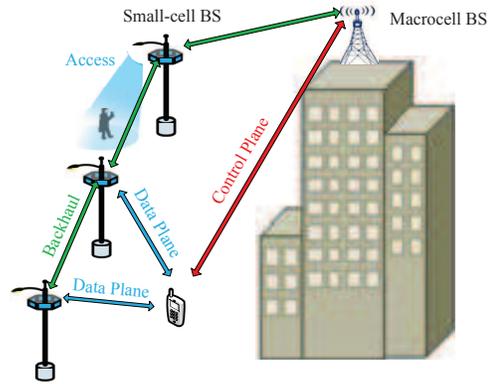}
\vspace*{-3mm}
\caption{MmWave Massive MIMO
based access and backhaul in UDN.
transmission.}
 \label{fig:HetNEt} 
\vspace*{-3mm}
\end{figure}

It has
been the consensus that future 5G networks
should achieve the 1000-fold increase in system
capacity~\cite{myWC,key,dong}. To realize such an aggressive 5G version,
millimeter-wave (mmWave) massive MIMO used for the access and backhaul
 in ultra-dense network (UDN) has been considered
as a promising technique to enable gigabit-per-second
user experience, seamless coverage, and green communication~\cite{myWC}.
In this paper, we consider the heterogeneous network (HetNet) with the separation of control plane and data plane, as shown in Fig. \ref{fig:HetNEt}.
 For such an HetNet, the macrocell base station (BS) provides the control signaling service for the
large coverage area using conventional low frequency band, while the ultra-dense small BSs are specialized in data resources with limited coverage area for high-rate transmission, where the emerging mmWave massive MIMO technique
 is exploited for the access and backhaul~\cite{myWC,in_band}. Moreover, the centralized radio access network (C-RAN)
 architecture can be integrated into the HetNet for the improved physical layer transmission, handover, scheduling,
 and etc., where the ultra-dense small cells are regarded as the remote ratio head (RRH) and macrocell BS is considered
 as the baseband unit (BBU).

Due to the ultra-dense deployment, each user may receive the signal from multiple RRHs.
To exploit the advantages of C-RAN architecture, the accurate channel state information (CSI)
associated with multiple RRHs known at BBU is essential for the
joint beamforming, scheduling, and cooperation among the ultra-dense
 small-cell BSs. However, how to require the reliable channel estimation with low
 overhead can be challenging, since the pilot overhead to estimate the
 channels associated with a large number of antennas in mmWave massive MIMO can be prohibitively high \cite{orthogonal}.
 \cite{Hur1,Hur2} have proposed a multilevel codebook based joint channel estimation and beamforming for
 mmWave access and backhaul. However, this scheme only considers the mmWave multi-antenna systems with
 analog beamforming, which is limited to the
 point-to-point based single-stream transmission. To solve this problem, the mmWave massive MIMO has been emerging \cite{Han_mag,myWC,Hea_JSTSP,Han_ref},
 where the hybrid analog-digital beamforming scheme is proposed
 to support the multi-stream transmission with low hardware cost and energy consumption.
 In \cite{Hea_JSTSP}, an adaptive channel estimation has been proposed for mmWave massive MIMO. However, this scheme
 is limited to the single-cell scenario. Besides, \cite{Han_ref} has proposed the reference signal design for
 the channel estimation in mmWave massive MIMO. However, the scheme fails to exploit the sparsity of mmWave channels, which may
 lead to a certain performance loss.

 Recent study and experiments have shown that the mmWave massive MIMO channels appear
 the obviously sparsity in the angular domain~\cite{myWC,Hea_JSTSP}, since
 the path loss for non-line-of-sight (NLOS) paths is much larger
than that for line-of-sight (LOS) paths in mmWave \cite{jiayi,ShenCL15}. Moreover, since the spatial
 propagation characteristics of the mmWave channels within the system bandwidth are nearly
 unchanged, such sparsity is shared by subchannels of different subcarriers when the widely
 used orthogonal frequency-division multiplexing (OFDM) is considered. This phenomenon is
 referred to as the \emph{spatially common sparsity within the system bandwidth} \cite{myCL2}.
%
 In this paper, we first propose the non-orthogonal pilot design at the small-cell BSs used for channel estimation.
 Furthermore, we propose a structured compressive sensing (SCS)-based sparse channel estimation scheme at the receiver
 to estimate the channels for the mmWave massive MIMO in UDN. 
Simulation results verify that the proposed scheme is superior to its
 counterpart, and it is capable of approaching the performance bound with low overhead.

 Throughout our discussions, the boldface lower and upper-case symbols denote column vectors and matrices, respectively. The Moore-Penrose inversion, transpose,
 and conjugate transpose operators are given by $(\cdot )^{\dag}$, $(\cdot )^{\rm T}$ and
 $(\cdot )^{*}$, respectively, while $\lceil \cdot \rceil$ is the integer ceiling operator. $(\cdot )^{-1}$ is the inverse operator. The $\ell_{0}$-norm and $\ell_{2}$-norm are
 given by $\|\cdot\|_0$ and $\|\cdot\|_2$, respectively, and $\left| \Gamma \right|$ is the
 cardinality of the set $\Gamma$. The support set of the vector $\mathbf{a}$ is denoted by
 ${\rm supp}\{\mathbf{a}\}$. The rank of ${\bf A}$ is denoted by ${\rm rank}\{{\bf A}\}$, while ${\rm E}\{\cdot \}$ is the
 expectation operator.
$\left( {\bf a} \right)_{\Gamma}$ denotes the entries of $\mathbf{a}$ whose
 indices are defined by $\Gamma$, while $\left( {\bf A}\right)_{\Gamma}$ denotes a
 sub-matrix of $\mathbf{A}$ with column indices defined by $\Gamma$. $ \otimes $ is the Kronecker product and ${\rm{vect}}\left( {\cdot} \right)$ is the vectorization operation according to the columns of the matrix. $\left[ {\bf a} \right]_i$ denotes the $i$th entry of the
 vector $\mathbf{a}$, and $\left[ {\bf A} \right]_{i,j}$ denotes the $i$th-row and
 $j$th-column element of the matrix $\mathbf{A}$.

\vspace*{-1mm}
\section{System Model}\label{S2}

Conventionally, mmWave is considered to be not suitable for access due to its high path loss,
Moreover, the signal-to-noise-ratio (SNR) before beamforming is conventionally considered to be low due to the high path loss,
which leads to the
challenging channel estimation in mmWave communications. However, for the UDN scenarios, we will clarify this
misunderstanding by comparing the path loss in UDN working at 30 GHz and that in conventional
cellular networks working at 3 GHz. Specifically, considering the multipath fading, signal dispersion, and other loss factors,
the path loss component of Friis equation in decibel (dB) can be provided as \cite{Hur1},
 \begin{equation}\label{equ:path} 
\eta  = 32.5 + 20{\log _{10}}\left( {{f_c}} \right) + 10\alpha {\log _{10}}\left( d \right) + \left( {{\alpha _o} + {\alpha _r}} \right)d,
\end{equation}
where $f_c$ (MHz) is the carrier frequency, $\alpha$ (dB/km) is the path loss exponent, $d$ (km) is the link distance, ${\alpha _o}$ (dB/km) is the atmospheric attenuation coefficient, and ${\alpha _r}$ (dB/km) is the rain attenuation coefficient. For the conventional cellular systems with $f_c=3$ GHz and $d=1$ km for example, we have $\eta =  192.62 $ dB, where $\alpha =2.2$ dB/km is considered in urban scenarios \cite{Hur1}, and the atmospheric attenuation and rain attenuation are ignored. By contrast, for UDN with $f_c=30$ GHz, we have $\eta =   188.27 $ dB with $d=100$ m for backhaul link and $\eta =    161.78 $ dB with $d=30$ m for access link, where $\alpha =2.2$ dB/km, $\alpha_o =0.1$ dB/km, and $\alpha_r =5$ dB/km when the heavy rain with 25 mm/h is considered \cite{key}. Due to the short link distance, the path loss in mmWave is even smaller than that in conventional cellular networks, which indicates the appropriate SNR for channel estimation in the mmWave access and backhaul even before beamforming.

\begin{figure}[!tp]
\centering
\includegraphics[width=6cm]{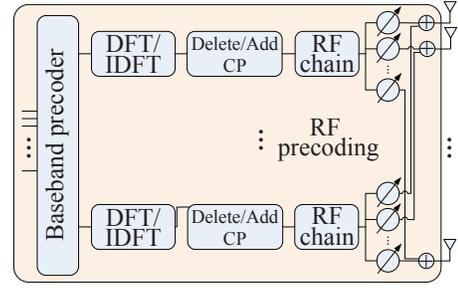}
\vspace*{-1mm}
\caption{Transceiver structure of the mmWave massive MIMO with analog phase shifter network.} \label{fig:Spectrum}
\vspace*{-5mm}
\end{figure}

On the other hand, mmWave massive MIMO has been emerging as the promising technique for access and backhaul \cite{Han_mag,myWC,Hea_JSTSP,Han_ref}. For the mmWave massive MIMO, as shown in Fig. \ref{fig:Spectrum}, the number of baseband chains is larger than one but far smaller than that of the employed antennas, and the hybrid analog and digital beamforming can be exploited for the improved spatial multiplexing with low hardware cost and energy consumption. However, the large number of antennas lead to the challenging issue of channel estimation. Particularly, the delay-domain mmWave massive MIMO channel can be modeled as~\cite{Hur1}\vspace*{-2mm}
\begin{equation}\label{equ0}
{{\bf{H}}^d}\left( \tau  \right) = \sum\nolimits_{l = 0}^{L - 1} {{\bf{H}}_l^d\delta \left( {\tau  - {\tau _l}} \right)} ,
\end{equation}
where $L$ is the number of multipath, $\tau_l$ is the delay of the $l$th path, $N_T$, $N_R$
are the numbers of antennas at the transmitter and receiver, respectively, and ${{\bf{  H}}^d_l} \in  {\mathbb{C}^{{N_R} \times {N_T}}}$ is given by
\begin{equation}
{\bf{ H}}^d_l{\rm{ = }}{\alpha _l}{{\bf{a}}_R}\left( {{\theta _l}} \right){\bf{a}}_T^*\left( {{\varphi _l}} \right),\label{equ1}
\end{equation}
with that ${{\alpha _l}}$ is the complex gain of the $l$th path, ${{\theta _l}}\in [0,2\pi]$ and ${{\varphi _l}}\in [0,2\pi]$
are azimuth angles of arrival or departure (AoA/AoD) if we consider the uniform linear array (ULA). 
In addition, ${{\bf{a}}_R}\left( {{\theta _l}} \right) = {\left[ {1,{e^{j2\pi d\sin ({\theta _l})/\lambda }}, \cdots ,{e^{j2\pi ({N_T} - 1)d\sin ({\theta _l})/\lambda }}} \right]^{\rm{T}}}$ and
${{\bf{a}}_T}\left( {{\varphi  _l}} \right) ={\left[ {1,{e^{j2\pi d\sin ({\varphi  _l})/\lambda }}, \cdots ,{e^{j2\pi ({N_R} - 1)d\sin ({\varphi  _l}) /\lambda }}} \right]^{\rm{T}}}$ are steering vectors at the receiver and transmitter, respectively,
where $\lambda $ and $d$ are wavelength and antenna spacing, respectively.

Since the path loss for NLOS paths is much larger than that for LOS paths in mmWave, the mmWave channels appear the obviously sparsity in the angular domain. Here we consider frequency-domain subchannel ${{\bf{ H}}^f_n}$ ($1\le n\le N$) at the $n$th
 subcarrier, where $N$ is the size of the OFDM symbol, and the relationship between the frequency-domain channel ${{\bf{ H}}^f_n}$ and the delay-domain channel ${\bf{ H}}^d(\tau)$ has been illustrated in our recent paper \cite{myCL2}. Moreover, we can obtain the sparse channel matrix in the angular domain ${{\bf{ H}}^a_n}$ as \cite{Hea_JSTSP}
 \begin{equation}
{{\bf{ H}}^f_n} = {{\bf{A}}_R}{{\bf{ H}}^a_n}{\bf{A}}_T^*,\label{equangular}
\end{equation}
where ${{\bf{A}}_T}\in \mathbb{C}^{N_T\times N_T}$ and ${{\bf{A}}_R}\in \mathbb{C}^{N_R\times N_R}$ are the discrete Fourier transformation (DFT) matrices by quantizing the virtual angular domain with the resolutions of $2\pi/N_T$ at the transmitter and $2\pi/N_R$ at the receiver, respectively. By vectorizing
${{\bf{ H}}^f_n}$, we can further obtain
 \begin{equation}
{{\bf{h}}_n^f} = {\rm{vect}}\left( {{{\bf{H}}^f_n}} \right) = \left( {({{\bf{A}}_T^*)}^{\rm{T}} \otimes {\bf{A}}_R} \right){\rm{vect}}\left( {{{{\bf{ H}}}^a_n}} \right) = {\bf{A}}{{\bf{ h}}^a_n},
\end{equation}
where ${\bf{A}}=\left( {({{\bf{A}}_T^*)}^{\rm{T}} \otimes {\bf{A}}_R} \right)$ and ${{\bf{ h}}^a_n}={\rm{vect}}\left( {{{{\bf{ H}}}^a_n}} \right) $.
 Due to the sparsity of ${\bf h}_n^a$, we can obtain that
\begin{equation}\label{equ:channelmode3} 
 \left|\Theta_n\right| = \left| {\rm supp}\left\{ {\bf h}_n^a \right\} \right| = S_a \ll N_T N_R ,
\end{equation}
 where $\Theta_n$ is the support set, and $S_a$ is the sparsity level in the angular domain. Note that if we consider
 the quantized AoA/AoD with the same resolutions as ${\bf{A}}_T$ and ${\bf{A}}_R$, we have $S_a=L$ \cite{Hea_JSTSP}.
 Since the spatial propagation characteristics of the channels within the system
 bandwidth are almost unchanged, $\left\{{\bf h}^a_n\right\}_{n=1}^{N}$
 have the common sparsity, namely,
 \begin{equation}\label{eq5}
 {\rm supp}\left\{{\bf h}^a_1\right\} = {\rm supp}\left\{{\bf h}_2^a\right\}
 = \cdots = {\rm supp}\left\{{\bf h}^a_N\right\} = \Theta ,
\end{equation}
which is referred to as the spatially common sparsity within the system bandwidth.

\vspace*{-2mm}

\section{SCS-Based Channel Estimation Scheme}\label{S3}
In this section, we propose the SCS-based channel estimation scheme for access in UDN, which can be also used to estimate the channels for backhaul. The procedure of the proposed channel estimation and the associated processing is first
 summarized. \emph{Step 1}: Under the control of the macrocell BS, several continuous time slots are specialized for
channel estimation, where ultra-dense small-cell BSs (RRH) transmit non-orthogonal pilots. \emph{Step 2}: The user uses the proposed SCS-based channel estimator to acquire the channels associated
with multiple small cells nearby.
\emph{Step 3}: The estimated channels are first quantized and then fed back to the macrocell (BBU)
for the centralized processing including the beamforming, scheduling, cooperation, and etc.

Due to the hybrid transceiver structure as shown in Fig. \ref{fig:Spectrum}, we consider each user employs $N_{a}^{\rm{US}}$ antennas and $N_{\rm BB}^{\rm{US}}$ baseband chains with
$N_{a}^{\rm{US}}\gg N_{\rm BB}^{\rm{US}}$, while each small-cell has $N_{a}^{\rm{BS}}$ antennas and
$N_{\rm BB}^{\rm{BS}}$ baseband chains with $N_{a}^{\rm{BS}}\gg N_{\rm BB}^{\rm{BS}}$.
Each user can receive the signal from $M$ small-cell BSs nearby.
In downlink channel estimation for one user,
the received pilot signal at the $\xi_p$th ($1 \le p \le P$) subcarrier
 in the $t$th time slot can be expressed as
\begin{equation}\label{equ:channelmode5}  
\!\!\!\!\!\!\begin{array}{l}
{\bf{r}}_p^{(t)} = {({\bf{Z}}_p^{(t)})^*}\sum\nolimits_{m = 1}^M {{{\bf{\tilde H}}_{p,m}^f}{\bf{f}}_{p,m}^{(t)}}  + {\bf{v}}_p^{(t)}\\
~~~~={({\bf{Z}}_p^{(t)})^*}\sum\nolimits_{m = 1}^M {{{\bf{A}}_R}{\bf{\tilde H}}_{p,m}^a{\bf{A}}_T^*{\bf{f}}_{p,m}^{(t)}}  + {\bf{v}}_p^{(t)},
\end{array}
\end{equation}
 where ${\bf{r}}_p^{(t)}\in \mathbb{C}^{N_{\rm{BB}}^{\rm US}\times 1}$ is the received signal dedicated to the $p$th pilot subcarrier in the $t$th
  time slot, ${\bf{Z}}_{p}^{(t)} = {\bf{Z}}_{{\rm{RF}},p}^{(t)}{\bf{Z}}_{{\rm{BB}},p}^{(t)}\in \mathbb{C}^{N_{a}^{\rm US}\times N_{\rm BB}^{\rm US}}$ is the combining matrix at the receiver with
  ${\bf{Z}}_{{\rm{BB}},p}^{(t)}\in \mathbb{C}^{N_{\rm BB}^{\rm US}\times N_{\rm BB}^{\rm US}}$ and ${\bf{Z}}_{{\rm{RF}},p}^{(t)}\in \mathbb{C}^{N_{a}^{\rm US}\times N_{\rm BB}^{\rm US}}$ the baseband and RF combining, respectively,
 ${{\bf{\tilde H}}_{p,m}^f} = {\bf{H}}_{{\xi _p},m}^f$ and ${{\bf{\tilde H}}_{p,m}}^a = {\bf{H}}_{{\xi _p},m}^a$ are frequency-domain and angular-domain channel matrices associated with the $p$th pilot subcarrier from the $m$th small-cell BS, respectively, $\Omega_{\xi}=\left\{ \xi_1,\xi_2,\cdots ,\xi_{P}\right\}$ is the index set of the
 pilot subcarriers, $\xi_p$ for $1 \le p \le P$ denotes the subcarrier index
 dedicated to the $p$th pilot subcarrier, ${\bf{f}}_{p,m}^{\left( t \right)} = {\bf{F}}_{{\rm{RF}},p}^{\left( t \right)}{\bf{F}}_{{\rm{BB}},p}^{\left( t \right)}{\bf{s}}_p^{\left( t \right)}\in \mathbb{C}^{N_{a}^{\rm BS}\times 1}$
 is the pilot signal transmitted by the $m$th small-cell BS, with
 ${\bf{F}}_{{\rm{RF}},p}^{\left( t \right)}\in \mathbb{C}^{N_{a}^{\rm BS}\times N_{\rm BB}^{\rm BS}}$,
  ${\bf{F}}_{{\rm{BB}},p}^{\left( t \right)}\in \mathbb{C}^{N_{\rm BB}^{\rm BS}\times N_{\rm BB}^{\rm BS}}$,
 ${\bf{s}}_p^{\left( t \right)}\in \mathbb{C}^{N_{\rm BB}^{\rm BS}\times 1}$ the RF precoding, baseband precoding, and
 training sequence, respectively,
 and ${\bf{v}}_p^{(t)}$ is the additive white Gaussian noise (AWGN) at the user. 
 Moreover, (\ref{equ:channelmode5}) can be simplified as
 \begin{equation}\label{equ:channelmode6}
\begin{array}{l}
{\bf{r}}_p^{(t)} = {({\bf{Z}}_p^{(t)})^*}{{\bf{A}}_R}{{{\bf{\bar H}}}^a_p}{\bf{\bar A}}_T^*{\bf{\bar f}}_p^{(t)} + {\bf{v}}_p^{(t)}\\
 = \left( {{{\left( {{\bf{\bar A}}_T^*{\bf{\bar f}}_p^{(t)}} \right)}^{\rm{T}}} \otimes {{({\bf{Z}}_p^{(t)})}^*}{{\bf{A}}_R}} \right){\rm{vect}}\left( {{{{\bf{\bar H}}}^a_p}} \right)
 = {{\bf{\Phi }}^{\left( t \right)}}{\bf{\bar h}}_p^{a},
\end{array}
\end{equation}
where ${\bf{\bar H}}_p^a = \left[ {{\bf{\tilde H}}_{p,1}^a,{\bf{\tilde H}}_{p,2}^a, \cdots ,{\bf{\tilde H}}_{p,M}^a} \right]\in \mathbb{C}^{N_{a}^{\rm US}\times MN_{a}^{\rm BS}}$, ${\bf{\bar A}}_T^* = {\rm{diag}}\left\{ {{\bf{A}}_T^*,{\bf{A}}_T^*, \cdots ,{\bf{A}}_T^*} \right\}\in \mathbb{C}^{MN_{a}^{\rm BS}\times MN_{a}^{\rm BS}}$, ${\bf{\bar f}}_p^{\left( t \right)} = {\left[ {{{\left( {{\bf{f}}_{p,1}^{(t)}} \right)}^{\rm{T}}},{{\left( {{\bf{f}}_{p,2}^{(t)}} \right)}^{\rm{T}}}, \cdots ,{{\left( {{\bf{f}}_{p,M}^{(t)}} \right)}^{\rm{T}}}} \right]^{\rm{T}}}\in \mathbb{C}^{MN_{a}^{\rm BS}\times 1}$, ${\bf{\bar h}}_p^a = {\rm{vect}}\left( {{\bf{\bar H}}_p^a} \right)\in \mathbb{C}^{MN_{a}^{\rm BS}N_{a}^{\rm US}\times 1}$,
 and ${{\bf{\Phi }}^{\left( t \right)}} = {\left( {{\bf{\tilde A}}_T^*{\bf{\tilde f}}_p^{(t)}} \right)^{\rm{T}}} \otimes {({\bf{Z}}_p^{(t)})^*}{{\bf{A}}_R}\in \mathbb{C}^{N_{\rm BB}^{\rm US} \times MN_{a}^{\rm BS}N_{a}^{\rm US}}$.
 Due to the quasi-static property of the channel within the coherence time, the received signals
 in $G$ successive time slots can be jointly exploited to acquire the downlink channel estimation
 at the user, which can be expressed as
\begin{align}\label{equ:joint_process2} 
{\bf{r}}_p^{[G]} = {\bf{\Phi }}_p^{[G]}\overline {\bf{h}} _p^a + {\bf{v}}_p^{[G]},
\end{align}
where ${\bf{r}}_p^{[G]} \!\!= \!\!{\left[ {{{\left( {{\bf{r}}_p^{(1)}} \right)}^{\rm{T}}},{{\left( {{\bf{r}}_p^{(2)}} \right)}^{\rm{T}}}, \cdots ,{{\left( {{\bf{r}}_p^{(G)}} \right)}^{\rm{T}}}} \right]^{\rm{T}}}\!\!\in \mathbb{C}^{GN_{\rm BB}^{\rm US}\times 1}$ is the aggregate received signal,
${{\bf{\Phi }}^{[G]}} \!\!\!\!= \!\!\!\!{\left[ {{{\left( {{{\bf{\Phi }}^{(1)}}} \right)}^{\rm{T}}},{{\left( {{{\bf{\Phi }}^{(2)}}} \right)}^{\rm{T}}}, \cdots ,{{\left( {{{\bf{\Phi }}^{(G)}}} \right)}^{\rm{T}}}} \right]^{\rm{T}}}\!\!\in\!\! \mathbb{C}^{GN_{\rm BB}^{\rm US} \times MN_{a}^{\rm BS}N_{a}^{\rm US}}$ is the aggregate measurement matrix,
and ${\bf{v}}_p^{[G]} \!\!= \!\!{\left[ {{{\left( {{\bf{v}}_p^{(1)}} \right)}^{\rm{T}}},{{\left( {{\bf{v}}_p^{(2)}} \right)}^{\rm{T}}}, \cdots ,{{\left( {{\bf{v}}_p^{(G)}} \right)}^{\rm{T}}}} \right]^{\rm{T}}}$ is aggregate AWGN. The system's SNR
 is defined as $\mbox{SNR}\!\!=\!\!{\rm E}\Big\{\Big\|{\bf \Phi}_p^{[G]}
 \bar{\bf h}_p^{a }\Big\|_2^2\Big\} \Big/ {\rm E}\Big\{\Big\| {\bf v}_p^{[G]}\Big\|_2^2\Big\}$,
 according to (\ref{equ:joint_process2}).

 To accurately estimate the
 channel from (\ref{equ:joint_process2}), the value of $G$ used in conventional algorithms,
 such as the minimum mean square error (MMSE) algorithm, is heavily dependent on the
 dimension of $\overline {\bf{h}} _p^a$, i.e., $MN_{a}^{\rm US}N_{a}^{\rm BS}$. Usually, $GN_{\rm BB}^{\rm US}\ge MN_{a}^{\rm US}N_{a}^{\rm BS}$ is required,
 which leads $G$ to be much larger than the coherence time and results in the
 poor channel estimation performance~\cite{orthogonal}. Moreover, to minimize the
 mean square error (MSE) of the channel estimate, ${\bf{\Phi }}_p^{[G]}$ should be a unitary
 matrix scaled by a transmit power factor \cite{orthogonal}. Usually, ${\bf{\Phi }}_p^{[G]}$
 is a diagonal matrix with equal-power diagonal elements. Such a pilot design is
 illustrated in Fig.~\ref{fig:pilot}\,(a), which is called the time-domain orthogonal
 pilot.
 It should be pointed out that in MIMO-OFDM systems, to estimate the channel
 associated with one transmit antenna, $P$ pilot subcarriers should be used, and usually
 $P=N_g$ is considered since $N_c=N/N_g$ adjacent subcarriers are correlated \cite{orthogonal}.
 Hence the total pilot overhead to estimate the complete MIMO channel can be as large as $P_{\rm total}=
 P G=N_g MN_{a}^{\rm US}N_{a}^{\rm BS}/N_{\rm BB}^{\rm US}$.
 In this paper, we will propose an efficient non-orthogonal pilot scheme.
\vspace*{-1mm}
\subsection{Non-Orthogonal Pilot for Downlink Channel Estimation}\label{non-orthogonal}
\begin{figure}[tp!]
\vspace*{-3mm}
\centering
\includegraphics[width=\columnwidth, keepaspectratio]{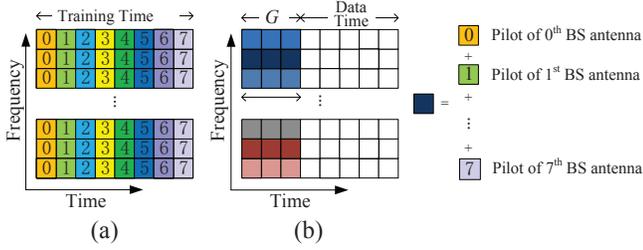}
\vspace*{-8mm}
\caption{(a)~Orthogonal pilot \cite{orthogonal}, (b)~Non-orthogonal pilot ($N_a^{\rm BS}=8$).}
 \label{fig:pilot} 
\vspace*{-5mm}
\end{figure}

 The proposed non-orthogonal pilot scheme is illustrated in Fig.~\ref{fig:pilot}\,(b).
 Similar to the time-domain orthogonal pilot scheme, $P$ subcarriers are dedicated to
 pilots in each OFDM symbol. However, the proposed scheme allows the non-orthogonal pilot
 signals associated with different BS antennas to occupy the completely identical
 frequency-domain subcarriers.

 The orthogonal pilot based conventional designs usually require $GN_{\rm BB}^{\rm US} \ge MN_{a}^{\rm US}N_{a}^{\rm BS}$. By contrast,
 the proposed non-orthogonal pilot for SCS-based channel estimator is capable of
 providing the efficient compression and reliable recovery of sparse signals. Therefore,
 $GN_{\rm BB}^{\rm US}$ is mainly determined by $S_a \ll MN_{a}^{\rm US}N_{a}^{\rm BS}/N_{\rm BB}^{\rm US}$. The non-orthogonal pilot of the first stage is
 designed in advance, which will be discussed in Section~\ref{S4.1}.
 For the placement of pilot subcarriers, the widely used equi-spaced pilot is considered. It is worth pointing out that the $p$th pilot
 subcarrier is shared by the pilot signals of the $N_a^{\rm BS}$ transmit antennas of $M$ small-cell BSs as illustrated
 in Fig.~\ref{fig:pilot}\,(b).

\vspace*{-1mm}
\subsection{SCS-Based Channel Estimation Scheme}\label{channel estimation}

 Given the measurements (\ref{equ:joint_process2}), the CSI can be acquired by solving the
 following optimization
\begin{align}\label{equ:target_func} 
\begin{array}{l}
 \min\limits_{\bar{\bf h}_p^{a }, 1\le p\le P} \Big( \sum\nolimits_{p=1}^{P}
 \left\| \bar{\bf h}_p^{a } \right\|_0^2 \Big)^{1/2} \\
 {\kern 15pt} {\rm s.t.} {\kern 2pt} {\bf r}_p^{[G]} = {\bf{\Phi}}_p^{[G]} \bar{\bf h}_p^{a },~{\forall p}
  \,~~{\rm and} ~ \left\{ \bar{\bf h}_p^{a } \right\}_{p=1}^{P}  \\ {\kern 30pt}
 \mbox{ share the common sparse support set.}
\end{array}
\end{align}
 To solve the optimization problem (\ref{equ:target_func}), we adopt the structured sparsity adaptive matching pursuit (SSAMP) algorithm to reconstruct the sparse angular domain
 channels of multiple pilot subcarriers \cite{TSP}.
 The SSAMP algorithm, listed in Algorithm~\ref{alg:Framwork}, is used to solve the
 optimization (\ref{equ:target_func}) to simultaneously acquire multiple sparse channel
 vectors at different pilot subcarriers. This algorithm is developed from the
 SAMP algorithm \cite{SAMP}.
 Compared to the classical SAMP algorithm \cite{SAMP} which recovers one high-dimensional
 sparse signal from single low-dimensional received signal, the SSAMP algorithm
 can simultaneously recover multiple high-dimensional sparse signals with the common
 support set by jointly processing multiple low-dimensional received signals.

 By using the SSAMP algorithm at the user, we can acquire $\widehat{\bar{\bf h}}_p^{a }$
 for $1\le p \le P$. Consequently, the actual $p$th pilot subchannel associated with the $m$th
  small-cell BS $\left\{ {{\bf{\tilde H}}_{p,m}^f} \right\}_{m = 1,p = 1}^{M,P}$ can be acquired.
  \begin{algorithm}[!tp]
{\small
\renewcommand{\algorithmicrequire}{\textbf{Input:}}
\renewcommand\algorithmicensure {\textbf{Output:} }
\caption{SSAMP Algorithm}
\label{alg:Framwork} 
\begin{algorithmic}[1]
\REQUIRE
 Noisy received signals ${\bf r}_p^{[G]}$ and sensing matrices ${\bf{\Phi}}_p^{[G]}$
 in (\ref{equ:target_func}), $1\le p \le P$; termination threshold $p_{\rm th}$.
\ENSURE
 Estimated channel vectors in the virtual angular domain at multiple pilot subcarriers
 $\widehat{\bar{\bf h}}_p^{a }$, $\forall p$.
\STATE ${\cal{T}} = 1$; $i = 1$; $j =1$. {\scriptsize \%~$\cal{T}$, $i$, $j$ are the sparsity
 level of the current stage, iteration index, and stage index, respectively.}
\STATE ${\bf c}_p\!=\!{\bf t}_p\! ={\bf c}_p^{\rm{last}}\!=\!{\bf{0}}\in \mathbb{C}^{M\times 1}$, $\forall p$.
 {\scriptsize \%~${\bf c}_p$ and ${\bf t}_p$ are intermediate variables, and ${\bf c}_p^{\rm{last}}$
 is the channel estimation of the last stage.}
\STATE $\Omega^0\!\!=\!\!\! \ {\Gamma}\!=\! \widetilde{\Gamma}\!=\!\Omega\!=\!\widetilde{\Omega}\!=\!
 \emptyset$; $l_{\text{min}}\!=\!\widetilde l\!=\!0$. {\scriptsize \%~$\Omega^i$ is the estimated
 support set in the $i$th iteration, $\Gamma$, $\widetilde{\Gamma}$, $\Omega$, and $\widetilde{\Omega}$
 are sets, $l_{\text{min}}$ and $\widetilde l$ denote the support indexes.}
\STATE ${\bf b}^0_p \!=\!{\bf r}_p^{[G]}\!\!\in \!\!\mathbb{C}^{G\times 1}\!$, $\forall p$.
 {\scriptsize \%~${\bf b}^i_p$ is the residual of the $i$th iteration.}
\STATE $\sum\nolimits_{p = 1}^P {\left\| {{\bf{b}}_p^{{\rm{last}}}} \right\|_2^2}=+\infty$.
 {\scriptsize \%~${{\bf{b}}_p^{{\rm{last}}}}$ is the residual of the last stage.}
\REPEAT
\STATE ${\bf a}_p=\left( {\bf{\Phi}}_p^{[G]} \right)^{*} {\bf b}^{i-1}_p$, $\forall p$.
 {\scriptsize \%~Signal proxy is saved in ${\bf a}_p$.}
\STATE $\Gamma \! = \!\arg \max\limits_{\widetilde{\Gamma}} \left\{ \! \sum\nolimits_{p=1}^{P} \!
 \left\| \left( {\bf a}_p \right)_{\widetilde{\Gamma}} \right\|_2^2 ,\left| \widetilde{\Gamma}
 \right| \! = \! {\cal{T}} \! \right\}$. {\scriptsize \%~Identify support.}
\STATE $\left( {\bf t}_p \right)_{ \Omega^{i-1} \cup \Gamma} \!\!=\!\!
 \left(\left( {\bf{\Phi}}_p^{[G]} \right)_{ \Omega^{i-1} \cup \Gamma}\right)^{\dag}
\!\!\! {\bf r}_p^{[G]}$, $\forall p$. {\scriptsize \%~LS estimation.}
\STATE $\Omega \! = \!\arg \max\limits_{\widetilde{\Omega}} \left\{ \! \sum\nolimits_{p=1}^{P} \!
 \left\| \left( {\bf t}_p \right)_{\widetilde{\Omega}} \right\|_2^2 ,
 \left| \widetilde{\Omega} \right| \! = \! {\cal T} \! \right\}$. {\scriptsize \%~Prune support.}
\STATE $\left( {\bf c}_p \right)_{\Omega} = \left(\left( {\bf{\Phi}}_p^{[G]} \right)_{\Omega}\right)^{\dag}
 {\bf r}_p^{[G]}$, $\forall p$. {\scriptsize \%~LS estimation.}
\STATE ${\bf b}_p = {\bf r}_p^{[G]} - {\bf{\Phi}}_p^{[G]} {\bf c}_p$, $\forall p$. {\scriptsize \%~Compute
 the residual.}
\STATE $l_{\rm min}\! = \! \arg \min\limits_{\widetilde l} \left\{ \sum\nolimits_{p=1}^{P}
 \left\| \left[ {\bf c}_p \right]_{\widetilde l} \right\|_2^2 , {\widetilde l} \in \Omega \right\}$.
\IF {$\sum\nolimits_{p=1}^{P} \left\| \left[ {\bf c}_p \right]_{l_{\min}}
 \right\|_2^2 /{P} < p_{\rm{th}}$}
    \STATE {Quit iteration.} 
\ELSIF{$\sum\nolimits_{p=1}^{P} \left\| {\bf b}_p^{\rm{last}} \right\|_2^2  < \sum\nolimits_{p=1}^{P}
 \left\| {\bf b}_p \right\|_2^2$}
    \STATE {Quit iteration.} 
\ELSIF{$\sum\nolimits_{p=1}^{P} \left\| {\bf b}_p^{i-1} \right\|_2^2 \le \sum\nolimits_{p=1}^{P}
 \left\| {\bf b}_p \right\|_2^2$}
     \STATE $j \!\!= \!\!j\!+\!1$; ${\cal{T}}\!\! = \!\!j$; ${\bf c}_p^{\rm{last}}
            \!\!=\!\! {\bf c}_p$, ${\bf b}_p^{\rm{last}}\! \!=\!\! {\bf b}_p$, $\forall p$.
\ELSE
    \STATE ${\Omega}^i = {\Omega}$; ${\bf b}_p^i = {\bf b}_p$, $\forall p$; $i  = i+1$.
\ENDIF
\UNTIL{$\sum\nolimits_{p=1}^{P} \left\| \left[ {\bf c}_p \right]_{l_{\min}} \right\|_2^2 / {P}
 < p_{\rm{th}}$}
\STATE $\widehat{\bar{\bf h}}_p^{a } = {\bf c}_p^{\rm{last}}$, $\forall p$. {\scriptsize \%~Obtain
 the final channel estimation.}
\end{algorithmic}
}
\end{algorithm}
\vspace*{-3mm}
\section{Performance Analysis}\label{S4}

 The performance analysis includes the non-orthogonal pilot design
 and the theoretical limit of the required time slot
 overhead for the SCS-based channel estimation scheme.

\vspace*{-1mm}
\subsection{Non-Orthogonal Pilot Design for Multi-Cell mmWave Massive MIMO Systems}\label{S4.1}

The measurement matrices ${\bf{\Phi}}_p^{[G]}$ $\forall p$ in
 (\ref{equ:joint_process2}) are very important for guaranteeing the reliable channel estimation
 at the user. Usually, $GN_{\rm BB}^{\rm US}\ll MN_{a}^{\rm US}N_{a}^{\rm BS}$. Since ${{\bf{\Phi }}^{[G]}} \!\!\!\!= \!\!\!\!{\left[ {{{\left( {{{\bf{\Phi }}^{(1)}}} \right)}^{\rm{T}}},{{\left( {{{\bf{\Phi }}^{(2)}}} \right)}^{\rm{T}}}, \cdots ,{{\left( {{{\bf{\Phi }}^{(G)}}} \right)}^{\rm{T}}}} \right]^{\rm{T}}}$,
  ${{\bf{\Phi }}^{\left( t \right)}} = {\left( {{\bf{\tilde A}}_T^*{\bf{\tilde f}}_p^{(t)}} \right)^{\rm{T}}} \otimes {({\bf{Z}}_p^{(t)})^*}{{\bf{A}}_R}$, ${\bf{\bar A}}_T^* = {\rm{diag}}\left\{ {{\bf{A}}_T^*,{\bf{A}}_T^*, \cdots ,{\bf{A}}_T^*} \right\}$, and ${\bf{ A}}_T$, ${\bf{ A}}_R$ are determined by the geometrical
  structure of
 the antenna arrays, both $\left\{ {{\bf{f}}_{p,m}^{(t)}} \right\}_{p = 1,m = 1,t=1}^{P,M,G}$ transmitted
 by the $M$ small-cell BSs and  $\left\{ {{{\bf{Z}}_p^{(t)}}} \right\}_{p = 1,t=1}^{P,G}$ at the user should be elaborated
 to guarantee the desired robust channel estimation.

 In CS theory, restricted isometry property (RIP) is used to evaluate the quality of the measurement matrix, in terms of
 the reliable compression and reconstruction of sparse signals. It is proven in \cite{STR_CS}
 that the measurement matrix with its elements following the independent and identically
 distributed (i.i.d.) complex Gaussian distributions satisfies the RIP and enjoys a satisfying
 performance in compressing and recovering sparse signals, which provides the viable pilot design guideline.

 On the other hand, the optimization problem (\ref{equ:target_func}) is essentially different from the
 single-measurement-vector (SMV) and multiple-measurement-vector (MMV) problems in CS\footnote{In CS, SMV recovers single high-dimensional sparse signal from its low-dimensional measurement signal, while MMV recovers multiple high-dimensional sparse signals with the common support set from multiple low-dimensional measurement signals with the identical measurement matrix.}.
 Typically, the MMV has the better recovery performance than the SMV, due to the potential
 diversity from multiple sparse signals \cite{STR_CS}. Intuitively, the recovery performance
 of multiple sparse signals with different measurement matrices, as defined in the generalized MMV (GMMV),
 should be better than that with the common measurement matrix as given in the MMV. This is
 because the further potential diversity can benefit from different measurement matrices for
 the GMMV. To prove this, we investigate the uniqueness of the solution to the GMMV problem.
 First, we introduce the concept of `spark' and the $\ell_{0}$-minimization based GMMV problem.

\begin{definition}\cite{STR_CS} \label{D1}
 The smallest number of columns of $\bf{\Phi}$ which are linearly dependent is the spark of
 the given matrix $\bf{\Phi}$, denoted by ${\rm spark}(\bf{\Phi})$.
\end{definition}

\begin{problem} \label{P1}
 $\min\limits_{{\bf x}_p,\forall p} \sum\limits_{p=1}^P \left\| {\bf x}_p \right\|_0^2 ,
~{\rm{s.t.}}~{\bf y}_p\!=\!{\bf{\Phi}}_p {\bf x}_p,~{\rm{supp}}\left\{ {\bf x}_p\right\}\!=\! \Xi,~\forall p$. 
\end{problem}

 For the above $\ell_{0}$-minimization based GMMV problem, ${\bf x}_p$, ${\bf y}_p$, ${\bf{\Phi}}_p$ are the high-dimensional sparse signal, low-dimensional measurement signal, and measurement matrix, respectively. Furthermore, the following result can be obtained.

\begin{theorem}\label{T1}
 \cite{TSP} For ${\bf{\Phi}}_p$, $1\le p\le P$, whose elements obey an i.i.d. continuous distribution,
 there exist full rank matrices ${\bf{\Psi}}_p$ for $2\le p\le P$ satisfying
 $\left( {\bf{\Phi}}_p \right)_{\Xi}={\bf{\Psi}}_p\left( {\bf{\Phi}}_1 \right)_{\Xi}$
 if we select $\left( {\bf{\Phi}}_1\right)_{\Xi}$ as the bridge, where $\Xi$ is the
 common support set. Consequently, ${\bf x}_p$ for $1\le p\le P$  will be the unique solution
 to Problem~\ref{P1} if
\begin{align}\label{equ:GMMV3} 
 2S <\rm{spark}\left( {\bf{\Phi}}_1 \right) - 1 + \rm{rank}\big\{ \widetilde{\bf Y}\big\} ,
\end{align}
 where $\widetilde{\bf Y}=\left[ {\bf y}_1 ~ {\bf{\Psi}}_2^{-1} {\bf y}_2 \cdots
 {\bf{\Psi}}_P^{-1} {\bf y}_P\right]$.
\end{theorem}

 From Theorem~\ref{T1}, it is clear that the achievable diversity gain introduced by
 diversifying measurement matrices and sparse vectors is determined by $\rm{rank}\big\{
 \widetilde{\bf Y}\big\}$. The larger $\rm{rank}\big\{ \widetilde{\bf Y}\big\}$ is, the
 more reliable recovery of sparse signals can be achieved. Hence, compared to the SMV and
 MMV, more reliable recovery performance can be achieved by the proposed GMMV. For the
 special case that multiple sparse signals are identical, the MMV reduces to the SMV
 since ${\rm rank}\left({\bf Y}\right)=1$, and there is no diversity gain by introducing
 multiple identical sparse signals. However, the GMMV in this case can still achieve
 diversity gain which comes from diversifying measurement matrices.

 According to the discussions above, a measurement matrix whose elements follow an i.i.d.
 Gaussian distribution satisfies the RIP. Furthermore, diversifying measurement matrices
 can further improve the recovery performance of sparse signals. This  enlightens us to
 appropriately design pilot signals.

 Specifically, as discussed above, we have ${\bf{Z}}_p^{(t)} = {\bf{Z}}_{{\rm{RF}},p}^{(t)}{\bf{Z}}_{{\rm{BB}},p}^{(t)}$,
   ${\bf{f}}_{p,m}^{\left( t \right)} = {\bf{F}}_{{\rm{RF}},p,m}^{\left( t \right)}{\bf{F}}_{{\rm{BB}},p,m}^{\left( t \right)}{\bf{s}}_{p,m}^{\left( t \right)}= {\bf{F}}_{{\rm{RF}},p,m}^{\left( t \right)}{\bf{\tilde s}}_{p,m}^{\left( t \right)}$ by defining ${\bf{\tilde s}}_{p,m}^{\left( t \right)}={\bf{F}}_{{\rm{BB}},p,m}^{\left( t \right)}{\bf{s}}_{p,m}^{\left( t \right)}$ for $1\le m \le M$ and $1 \le t \le G$, and $1\le p \le P$, each element of pilot signals is given by
\begin{align} 
&{\left[ {{\bf{Z}}_{{\rm{RF}},p}^{(t)}} \right]_{i_1,j_1}}\!\!\!\! \!= \!{e^{j{\phi^1_{i_1,j_1,t}}}},1 \le i_1 \le N_a^{{\rm{US}}},{\mkern 1mu} 1 \le j_1 \le N_{{\rm{BB}}}^{{\rm{US}}},\\ 
&{\left[ {{\bf{F}}_{{\rm{RF}},p}^{\left( t \right)}} \right]_{i_2,j_2}} \!\!\!\! \!= \! {e^{j{\phi^2 _{i_2,j_2,t,m}}}},1 \le i_2 \le N_a^{{\rm{BS}}},{\mkern 1mu}  \le j_2 \le N_{{\rm{BB}}}^{{\rm{BS}}},\\ 
&{\left[ {{\bf{s}}_p^{\left( t \right)}} \right]_{i_3}} = {e^{j{\phi^3_{i_3,p,t,m}}}},1 \le i_3 \le N_{{\rm{BB}}}^{{\rm{BS}}}, \\ 
&{\left[ {{\bf{Z}}_{{\rm{BB}},p}^{(t)}} \right]_{i_4,j_4}} \!\!\!\! \!= \! {e^{j{\phi^4 _{i_4,j_4,p,t}}}},1 \le i_4 \le N_a^{{\rm{BS}}},{\mkern 1mu} 1 \le j \le N_{{\rm{BB}}}^{{\rm{BS}}},
\end{align}
 where ${\phi^1_{i_1,j_1,t}}$, ${\phi^2 _{i_2,j_2,t,m}}$, ${\phi^3 _{i_3,p,t,m}}$, and ${\phi^4_{i_4,j_4,p,t}}$ follow the
 i.i.d. uniform distribution ${\cal{U}}\left[ 0, ~ 2\pi\right)$. Note that ${{\bf{F}}_{{\rm{RF}},1}^{\left( t \right)}}={{\bf{F}}_{{\rm{RF}},p}^{\left( t \right)}}$ and ${{\bf{Z}}_{{\rm{RF}},1}^{\left( t \right)}}={{\bf{Z}}_{{\rm{RF}},p}^{\left( t \right)}}$, since different subcarriers share the same RF precoding/combining. It is readily seen that
 the designed pilot signals guarantee that the elements of
 ${\bf{\Phi}}_p^{[G]}$,  obey the i.i.d. complex
 Gaussian distribution with zero mean.
Moreover, ${\bf{\Phi}}_p^{[G]}$ with different $p$ are diversified.
 Hence, the proposed pilot signal design is `optimal', in terms of the reliable compression
 and recovery of sparse angular domain channels.

\vspace*{-1mm}
\subsection{Required Time Slot Overhead for SCS-Based Channel Estimation}\label{S4.2}
 According to Theorem~\ref{T1}, for the optimization problem (\ref{equ:target_func}),
 $\widetilde{\bf Y}\!=\!{\bf{\Phi}}_1^{[G]} {\bf X}$ with
 $\widetilde{\bf Y}\!=\!\big[ {\bf r}_1^{[G]} ~ {\bf{\Psi}}_2^{-1} {\bf r}_2^{[G]}$ $\cdots
 {\bf{\Psi}}_{P}^{-1} {\bf r}_{P}^{[G]}\big]$ and ${\bf X}\!\!=\!\!\left[ \bar{\bf h}_1^{a } ~
 \bar{\bf h}_2^{ a} \cdots \bar{\bf h}_{P}^{ a} \!\right]$. Since $\big| {\rm supp}\big\{\bar{\bf h}_p^{a }
 \big\}\big|=S_a$, it is clear that
\begin{align}\label{equ:T_overhead1} 
 {\rm{rank}}\big\{ \widetilde{\bf Y} \big\} \le {\rm{rank}}\left\{ {\bf X} \right\}\le S_a .
\end{align}
 Moreover, as ${\bf{\Phi}}_1^{[G]}\in \mathbb{C}^{GN_{\rm BB}^{\rm US} \times MN_{a}^{\rm BS}N_{a}^{\rm US}}$,
\begin{align}\label{equ:T_overhead2}  
 {\rm{spark}}\big( {\bf{\Phi}}_1^{[G]} \big) \in \left\{2,3,\cdots ,GN_{\rm BB}^{\rm US}+1\right\} .
\end{align}
 Substituting (\ref{equ:T_overhead1}) and (\ref{equ:T_overhead2}) into (\ref{equ:GMMV3})
 yields $GN_{\rm BB}^{\rm US}\ge S_a+1$. Therefore, the smallest required time slot overhead is $G = \left\lceil {\frac{{{S_a} + 1}}{{N_{{\rm{BB}}}^{{\rm{US}}}}}} \right\rceil $.
  By increasing the number of measurement vectors $P$, the required time slot overhead
 for reliable channel estimation can be reduced, since more measurement matrices
 and sparse signals can increase ${\rm{rank}}\big\{  \widetilde{\bf Y}\big\}$.

\vspace*{-2mm}
\section{Simulation Results}\label{S5}

 We consider the ULA-based mmWave massive MIMO system with $d ={\lambda}/{2}$.
 In the simulations, $f_c=30$\,GHz, bandwidth $B_s=0.25$\,GHz, $N_a^{\rm US}=32$,
 $N_{\rm BB}^{\rm US}=2$, $N_{a}^{\rm BS}=512$, $N_{\rm BB}^{\rm BS}=8$, $\tau_{\rm max}=100$ ns, $N=P=64>\tau_{\rm max}B_s$,
 the user
 simultaneously estimates the channels associated with $M=4$ small-cell BSs nearby,
 $L=4$ for each link between the user and small-cell BS. For mmWave massive MIMO channels,
 we consider Rican fading consisting of one LOS path and $L-1$ equal-power NLOS paths with $K_{\rm{factor}}=10$ dB, where path gains follow the mutually independent complex Gaussian distribution with zero means, and $K_{\rm{factor}}$ denotes the ratio between the power of LOS path and the power of NLOS path. We set $p_{\rm{th}}$ to 0.06, 0.02, 0.01, 0.008, and 0.005,
 respectively, at the SNR of 10\,dB, 15\,dB, 20\,dB,
 25\,dB and $\ge 30$\,dB.  The oracle LS estimator was used as the benchmarks
 for the SCS-based channel estimation scheme.
 The adaptive orthogonal matching pursuit (OMP)-based channel estimation scheme \cite{Hea_JSTSP} was also adopted for comparison.

\begin{figure}[tp!]
\vspace*{-7mm}
\begin{center}
\includegraphics[width=0.85\columnwidth, keepaspectratio]{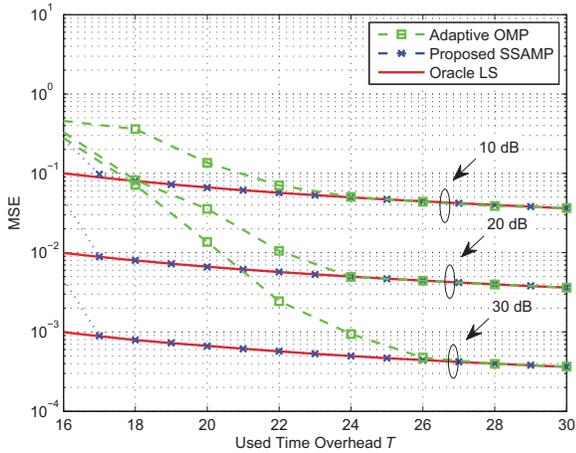}
\end{center}
\vspace*{-5mm}
\caption{MSE performance of different channel estimation schemes as functions of the time
 overhead $G$ and SNR.}
\label{fig:mse_vs_T} 
\vspace*{-4mm}
\end{figure}

 Fig.~\ref{fig:mse_vs_T} compares the MSE performance of the adaptive OMP scheme and the SSAMP algorithm,
 where $S_a=LM=16$ was considered. The oracle LS
 estimator with the known support set of the sparse channel vectors was adopted as the
 performance bound. From Fig.~\ref{fig:mse_vs_T}, it can be seen that the adaptive OMP scheme performs poorly.
 By contrast, the SSAMP algorithm is capable
 of approaching the oracle LS performance bound when $G >2S_a/N_{\rm BB}^{\rm US}$. This is because the proposed
 SCS-based scheme fully exploits the spatially common sparsity of mmWave channels within the system bandwidth.

\begin{figure}[!t]
\begin{center}
\vspace*{-7mm}
\includegraphics[width=0.85\columnwidth, keepaspectratio]{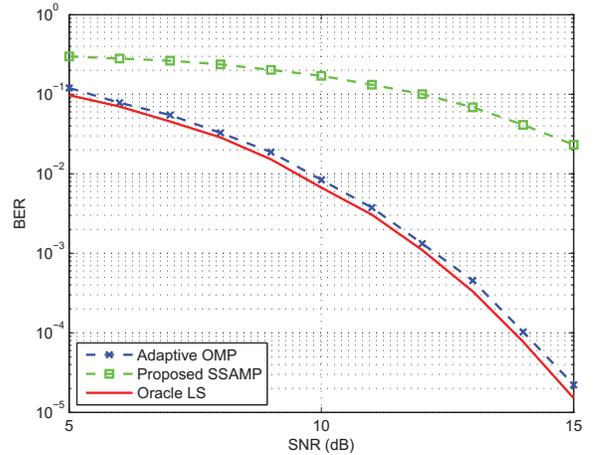}
\end{center}
\vspace*{-5mm}
\caption{Downlink BER performance, where the used channel state information at the BS is
 acquired by different channel estimation schemes.}
\label{fig:ber_vs_snr} 
\vspace*{-4mm}
\end{figure}

 Fig.~\ref{fig:ber_vs_snr} compares the downlink bit error rate (BER) performance
 by using the hybrid analog-digital beamforming proposed in our previous work \cite{myWC}, where 16 QAM is used, and the precoding and combining are
  based on the estimated channels\footnote{We consider the estimated channels are perfectly fed back to the macrocell BS (BBU) with the control plane using low frequency band. The feedback overhead can be small, since only the estimated support set and the associated path gains are enough due to the sparsity of mmWave channels.}. In the simulations, we consider two best LOS paths are used to serve the user due to $N_{\rm BB}^{\rm US}=2$, which indicates two out of four small-cell BSs with the optimal channel quality jointly serve the user. It can be
 observed that the proposed channel estimation scheme outperforms its counterpart,
 and its BER performance is capable of approaching that of the performance bound.

\vspace*{-2mm}
\section{Conclusions}\label{S6}

In this paper, we have proposed the SCS-based channel estimation scheme for the
mmWave massive MIMO based access and backhaul in UDN.
We first demonstrate that the SNR before beamforming in mmWave is appreciate for channel estimation
due to the short link distance in UDN, although the path loss in mmWave is high. Moreover, by exploiting
the sparsity of mmWave channels in the angular domain due to the high path loss for NLOS paths in mmWave,
we propose the non-orthogonal pilot at the transmitter and the SCS-based channel estimator at the receiver. The proposed
scheme can simultaneously estimate the channels associated with multiple small-cell BSs, and the required
pilot overhead is only dependent on the small number of the dominated multipath.
Simulation results
 have confirmed that our scheme can reliably acquire the mmWave massive MIMO channels with much reduced pilot overhead.
\section{Acknowledgments}
{This work was supported in part by the International Science \& Technology Cooperation Program of China (Grant No. 2015DFG12760),
the National Natural Science Foundation of China (Grant Nos. 61571270 and 61271266),  the Beijing Natural Science Foundation (Grant No. 4142027),
and the Foundation of Shenzhen government.}

\vspace*{-15mm}

\end{document}